# ARTICLE INFORMATION

**Article title**

*Multi-mode Fault Diagnosis Datasets of Three-phase Asynchronous Motor Under Variable Working Conditions*

**Authors**

*Shijin Chen[a,b], Zeyi Liu[c,d], Chenyang Li[e], Dongliang Zou[a], Xiao He[c,d]*, Donghua Zhou[c,f]*

**Affiliations**

*a MCC5 Group Shanghai Co. LTD, 201900, ShangHai, China.*

*b Shanghai Bearing Technology Research Institute Co. LTD, Shanghai, 201801, China.*

*c Department of Automation, Tsinghua University, 100084, Beijing, China.*

*d Institute for Embodied Intelligence and Robotics, Tsinghua University, 100084, Beijing, China.*

*e School of Mechanical and Power Engineering, Zhengzhou University, 450001, Zhengzhou, China.*

*f School of Automation, Southeast University, 210096, Nanjing, China.*

**Corresponding author's email address and Twitter handle**

*hexiao@tsinghua.edu.cn*

**Abstract**

Three-phase asynchronous motor are fundamental components in industrial systems, and their failure can lead to significant operational downtime and economic losses. Vibration and current signals are effective indicators for monitoring motor health and diagnosing faults. However, motors in real applications often operate under variable conditions such as fluctuating speeds and loads, which complicate the fault diagnosis process. This paper presents a comprehensive dataset collected from a three-phase asynchronous motor under various fault types and severities, operating under diverse speed and load conditions. The dataset includes both single faults and mechanical-electrical compound faults, such as rotor unbalance, stator winding short circuits, bearing faults, and their combinations. Data were acquired under both steady and transitional conditions, with signals including triaxial vibration, three-phase currents, torque, and key-phase signals. This dataset supports the development and validation of robust fault diagnosis methods for electric motors under realistic operating conditions.

## SPECIFICATIONS TABLE

| | |
|---|---|
| **Subject** | *Mechanical engineering* |
| **Specific subject area** | *Machine condition monitoring, motor fault diagnosis* |
| **Type of data** | *Datasets in ".csv" format* |
| **Data collection** | *Eight-channel synchronous vibration–current–torque–keyphase data were recorded from a 2.2 kW three-phase induction motor under variable speed/load profiles at 12.8 kHz. Controlled single and compound faults were physically introduced and measured over 282 runs (90 s each), saved as CSV files.* |
| **Data source location** | *MCC5 Group Shanghai Co. LTD*<br>*Department of Automation, Tsinghua University* |
| **Data accessibility** | **Repository name:** *Multi-mode Fault Diagnosis Datasets of Three-phase Asynchronous Motor Under Variable Working Conditions*<br>**Data identification number:** 10.17632/6s3dggj9mw.1<br>**Direct URL to data:** https://data.mendeley.com/datasets/6s3dggj9mw/1<br>**Instructions for accessing these data:** |
| **Related research article** | Z. Liu, C. Li and X. He, Evidential ensemble preference-guided learning approach for real-time multimode fault diagnosis, IEEE Transactions on Industrial Informatics, doi: 10.1109/TII.2023.3332112. |

## VALUE OF THE DATA

1) Unlike most publicly available motor fault datasets that focus on single fault types under steady states, this dataset systematically includes 24 fault conditions encompassing not only common single electrical and mechanical faults but also deliberately designed mechanical–electrical compound faults. These data were collected under 12 speed and load profiles, including both steady and transitional conditions, thereby supporting the development and evaluation of diagnostic methods that must perform in real-world, time-varying industrial environments.
2) The simultaneous recording of triaxial vibration, three-phase currents, torque, and key-phase signals at 12.8 kHz provides a coherent multimodal perspective on fault evolution. Researchers can exploit the complementary nature of vibration (sensitive to mechanical impacts) and current (sensitive to electromagnetic anomalies) signals to develop fusion-based or cross-domain diagnostic models, which are promising for improving the accuracy and robustness of fault isolation, especially for compound faults.

3) The dataset is structured with clearly defined operating condition variations and fault severity levels, allowing systematic investigation into domain adaptation, transfer learning, and generalization of data-driven models across different working conditions. This addresses a key challenge in industrial applications where models trained under one condition often degrade when applied to another.
4) With well-documented fault types, severity levels, and condition labels, the dataset is suitable for validating classical signal processing and machine learning methods. At the same time, the raw, high-sample-rate, multi-channel time-series format allows researchers to design and test deep neural networks for raw signal classification, feature learning, and real-time diagnosis.
5) All data are publicly accessible in a standardized CSV format with detailed metadata describing each test run. This transparency enables other researchers to replicate studies, compare algorithm performance under consistent conditions, and accelerate progress in motor fault diagnosis research.

Table 1 Comparison of several representative datasets

| | Taiyuan University of Technology [1] | Huazhong University of Science and Technology [2] | Korea Advanced Institute of Science and Technology [3] | MCC5-THU motor fault diagnosis datasets |
|---|---|---|---|---|
| Number of fault types | 6 | 6 | 2 | 24 |
| Number of signals | 9 | 3 | 4 | 8 |
| Sampling frequency | / | 25.6 kHz | 25.6 kHz | 12.8 kHz |
| Sampling period | / | / | 120 s | 90 s |
| Number of steady conditions | / | 2 | 1 | 24 |
| Key variables | / | Speed | / | Speed, Load |
| Number of transitional conditions | 0 | / | 0 | 48 |
| Number of compound faults | 1 | / | 0 | 9 |
| Number of fault degrees of severity | 3 | / | >10 | 2 |

## BACKGROUND

As a core component of electrically driven systems, the asynchronous motor is widely employed in industrial and mining enterprises due to its straightforward control and low maintenance requirements, serving as a prime mover for various types of machinery and industrial equipment. The motor is composed of the stator, rotor, air gap, and bearings, among other components, where the stator generates a rotating magnetic field and the rotor induces current to produce torque. Although its structure is relatively simple, harsh operating conditions, such as heavy loads, challenging environments, and complex electromagnetic interactions, often lead to incipient faults during operation[4,5]. Common failures include inter-turn short circuits in stator windings, broken rotor bars, bearing damage, air gap eccentricity, and rotor imbalance[6-7].

In motor fault diagnosis, stator current signals are widely adopted for online detection of rotor bar breakages due to their easy acquisition and robustness to external disturbances. Meanwhile, vibration signals collected from bearings via sensors allow for the extraction of characteristic fault-related frequencies, representing one of the most direct and effective approaches for diagnosing mechanical faults in bearings[8-10]. As an exemplary electromechanical device, motor faults can originate from either electrical or mechanical components. In industrial practice, distinguishing between mechanical and electrical faults is often challenging due to their similar manifestations, which significantly hinders the efficiency of on-site diagnosis. Therefore, establishing a comprehensive motor fault dataset that incorporates multi-source information such as vibration and current signals is crucial. Analysing motor conditions from both electrical and mechanical perspectives can substantially improve the accuracy and efficiency of fault diagnosis.

# DATA DESCRIPTION

The experimental setup, shown in Fig. 1, consists of a 2.2 kW three-phase asynchronous motor, a torque sensor, a two-stage parallel gearbox (with planetary gearbox removed), a magnetic powder brake serving as the load, and a measurement and control system. Faults were introduced by replacing bearings at the drive end and fan end of the motor, simulating single and compound faults under various operating conditions. Detailed motor and bearing specifications are provided in Table 2 and Table 3, respectively.

Table 2 Motor parameters

| Parameter | Value | Parameter | Value |
|---|---|---|---|
| Motor pattern | QABP-90L2 | Stator inner diameter | 72 mm |
| Poles of pair | 2 | Stator outer diameter | 130 mm |
| Single-side air gap length | 0.5 mm | Rotor inner diameter | 25 mm |
| Bearings | 6205.2Z-C3/6205.2Z-C3 | Rotor outer diameter | 71 mm |

Table 3 Bearing parameters

| Parameter | Value |
|---|---|
| Pitch diameter | 39.04mm |
| Ball diameter | 7.94mm |

| Ball number | 9 |
|---|---|
| Contact angle | 39.04mm |

A total of 282 test runs were conducted under 12 operating conditions, covering both steady and transitional states. Load-time and speed-time profiles are illustrated in Fig. 2(a) and Fig. 2(b), respectively. The dataset includes 24 fault types, encompassing healthy state, rotor imbalance, bending, broken bars, stator winding short circuits (low and high severity), voltage unbalance, static and dynamic eccentricity, bearing faults (inner, outer, ball, and compound), and mechanical-electrical compound faults.

Data were acquired at a sampling frequency of 12.8 kHz through 8 synchronous channels: key-phase, torque (measured using torque sensor model S2001, ±0.5 %F.S accuracy, sensitivity 100 mV/Nm), triaxial motor vibration (measured using three-axis vibration acceleration sensors model TES001V, sensitivity 100 mV/g), and three-phase currents (measured using current clamps model Fluke-i30s, sensitivity 100 mV/A). All recorded data represent raw voltage values, which can be converted into the corresponding physical quantities using the sensitivity coefficients provided above. Each data file contains 9 columns, and the meaning of each column of data is as shown in Table 4.

Table 4 The meaning of each column in the dataset

| column | Meaning |
|---|---|
| First column | Sampling time |
| Second column | Motor key-phase signal, from which the rotational speed can be derived |
| Third column | Torque on the gearbox input shaft |
| Fourth column | Vibration acceleration at the motor drive end in the horizontal radial direction |
| Fifth column | Vibration acceleration at the motor drive end in the axial direction |
| Sixth column | Vibration acceleration at the motor drive end in the vertical radial direction |
| Seventh column | Phase A current of the motor |
| Eighth column | Phase B current of the motor |
| Ninth column | Phase C current of the motor |

A subset of filenames is provided in Table 5 to illustrate the corresponding operating conditions and fault types.

Table 5 Example description of a partial dataset file.

| Filename | Description |
|---|---|
| Bearing_inner_L_speed_circulation_20Nm_1000rpm | The dataset was acquired from a motor with a light-severity inner raceway defect (length: 0.2 mm, depth: 0.5 mm), operating at a constant output torque of 20 Nm and following the 0–1000 rpm speed-time profile in Fig. 2(b). |
| Bearing_inner_L_torque_circulation_20Nm_1000rpm | The dataset was acquired from a motor with a light-severity inner raceway defect (length: 0.2 mm, depth: 0.5 mm), operating at the 1000 rpm. The torque-time curve 0-20NM shown in Fig. 2 (a). |
| Bearing_inner_H_speed_circulation_20Nm_1000rpm | A motor dataset for high-severity inner raceway fracture. Single fracture length is 0.6mm and depth is 0.5mm. Motor output shaft torque is 20 Nm. The motor shaft rotates at the 0–1000 rpm speed-time curve shown in Fig. 2 (b). |
| Bearing_inner_H_torque_circulation_20Nm_1000rpm | A motor dataset for high-severity inner raceway fracture. Single fracture length is 0.6mm and depth is 0.5mm. Motor shaft rotates at the 1000 rpm. The torque-time curve 0-20NM shown in Fig. 2 (a). |
| Winding_H_and_bearing_outer_H_speed_circulation_20Nm_1000rpm | A motor dataset for high winding short circuit and high outer raceway fracture. The winding fault is a 10% short circuit fault in a single winding. Single fracture length is 0.6mm and depth is 0.5mm. Motor shaft rotates at the 1000 rpm. The torque-time curve 0-20NM shown in Fig. 2(a). |
| Winding_H_and_bearing_inner_H_torque_circulation_40Nm_3000rpm | A motor dataset for high winding short circuit and high inner raceway fracture. The winding fault is a 10% short circuit fault in a single winding. Single fracture length is 0.6mm and depth is 0.5mm. Motor shaft rotates at the 3000 rpm. The torque-time curve 0-40NM shown in Fig. 2(a). |

# EXPERIMENTAL DESIGN, MATERIALS AND METHODS

The test rig (Fig. 1) comprises a 2.2 kW three-phase asynchronous motor, a torque sensor  a two-stage parallel gearbox, and a magnetic powder brake serving as the load. A triaxial vibration acceleration sensor was mounted on the motor drive end. Vibration, three-phase current, torque, and key-phase signals were synchronously acquired at a sampling frequency of 12.8 kHz using a multi-channel data acquisition system. This frequency was chosen as it according to the Nyquist sampling theorem is more than twice the highest characteristic frequency of all targeted motor faults (e.g., bearing defect frequencies and current harmonics), while also maintaining a manageable data volume for the multi-channel, long-duration recordings. The motor was tested under 12 operating conditions, including constant speed with variable load and constant load with variable speed. The laboratory temperature was controlled within ±2°C to minimize experimental error. Faults were introduced via laser etching with a precision of 0.01 mm. The dataset includes 282 test recordings, each 90 seconds long, stored in CSV format.

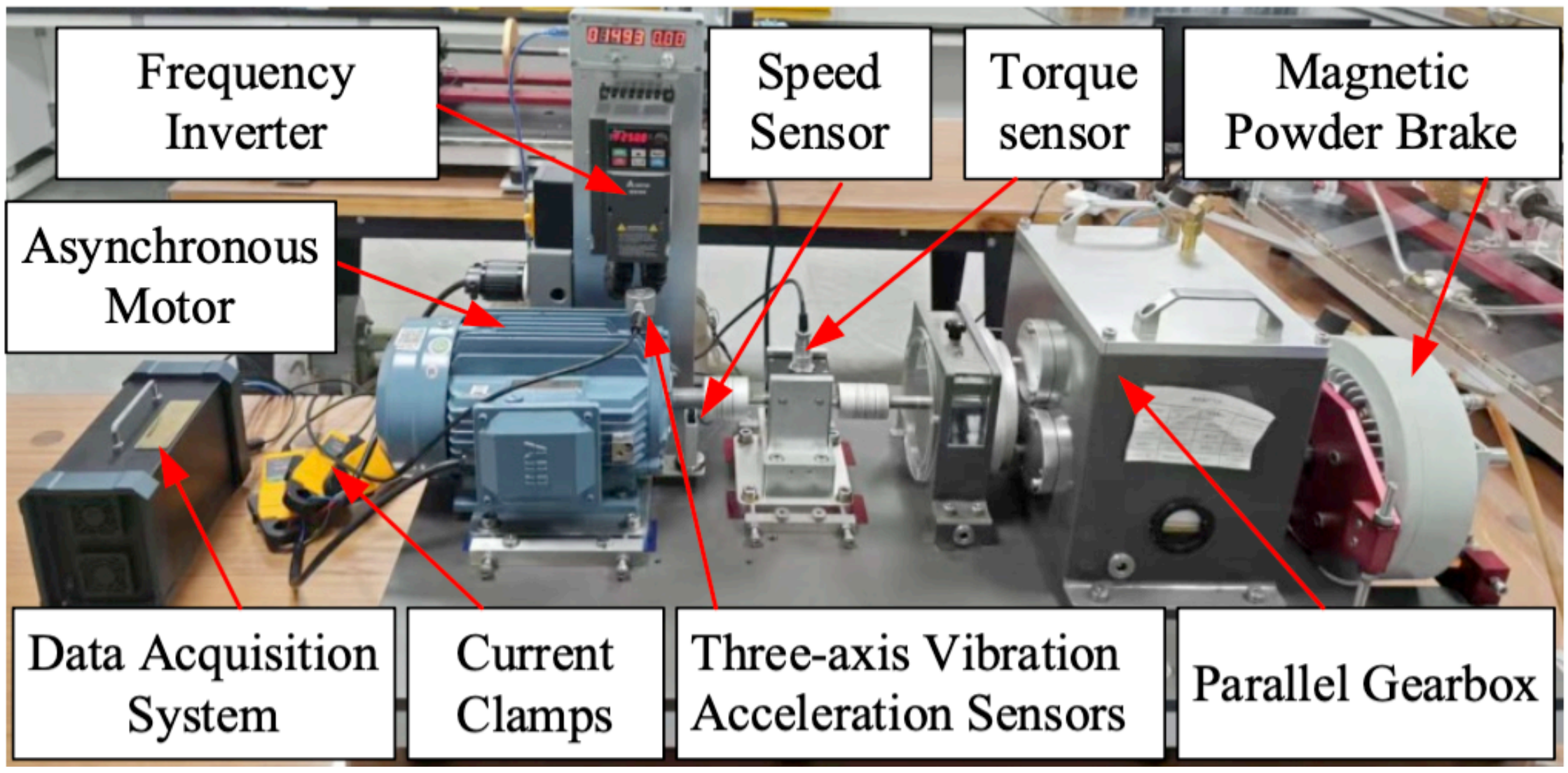

Figure 1 The actual motor test rig

Two load scenarios were implemented: constant-speed variable-torque and constant-torque variable-speed operations, as illustrated in Figure 2(a) and Figure 2(b), respectively. For constant-speed variable-torque cases (Fig. 2a), motor speed was held constant at 1000, 2000, or 3000 rpm. For example, under the 0-15-20 Nm condition, torque was set to 20 Nm during 10-30 s and 60-80 s intervals, and to 15 Nm during 35-55 s. For constant-torque variable-speed cases, load was maintained at 20 Nm or 40 Nm while speed followed the profile in Fig. 2(b). Due to magnetic hysteresis in the motor and torque generator, actual speed-torque-time profiles exhibited slight deviations from preset trajectories.

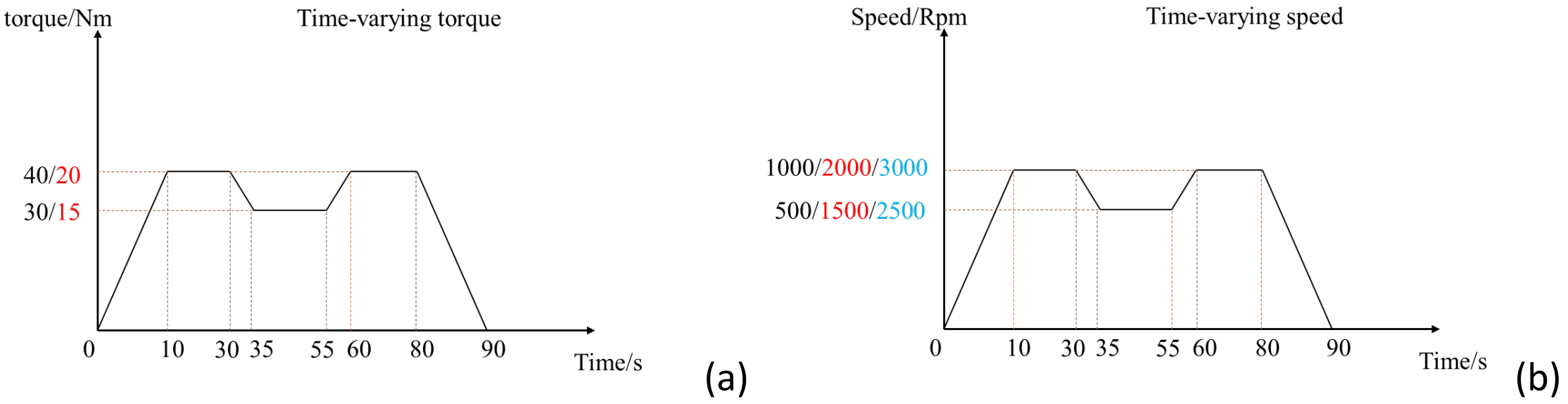

Figure 2 Time-varying operating condition curve (a) Time-varying torque curve (b) Time-varying speed curve

To simulate inter-turn stator winding faults caused by insulation degradation, a variable shunt resistor was connected in parallel across two taps of the target phase winding, as illustrated in Figure 3. The equivalent shunt resistance, Rf, could be adjusted from infinity (healthy condition) to near zero (severe short circuit). Two distinct fault severity levels were achieved by tuning Rf to induce circulating currents equivalent to 5% and 10% of the rated phase current, respectively. These levels were subsequently used to quantify fault-related signatures.

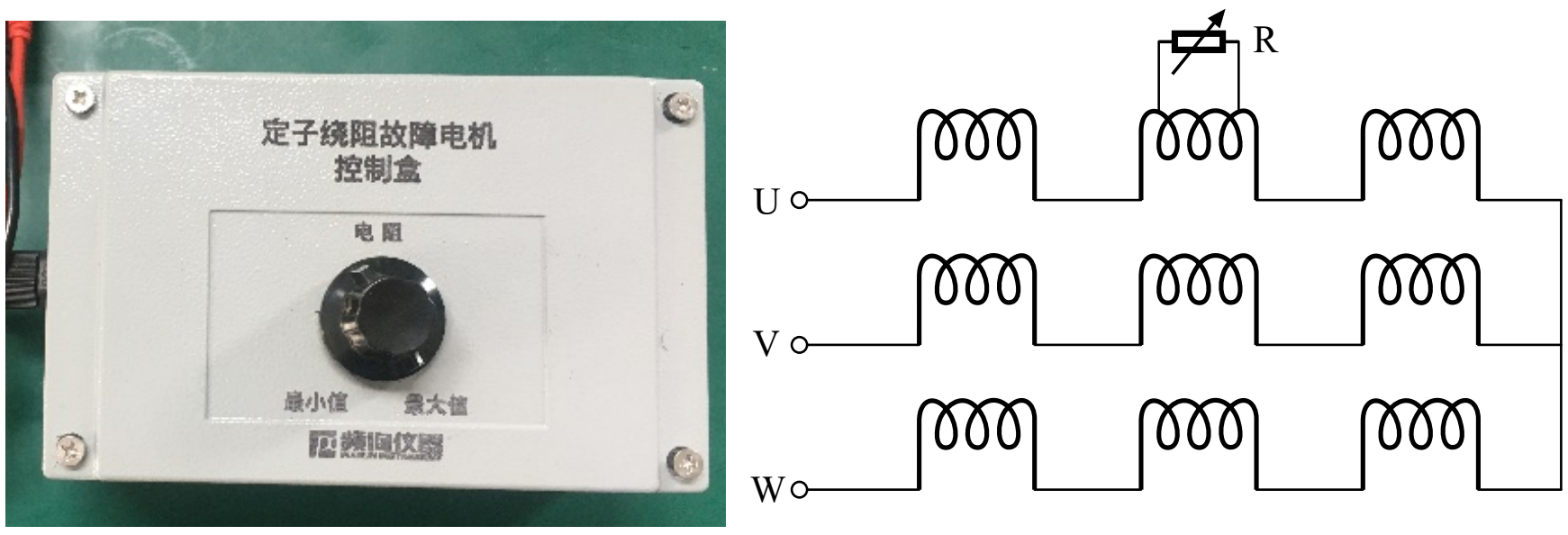

Fig.3. Stator winding fault controller.

Static radial eccentricity was introduced by symmetrically installing two fine-adjustment screws on the motor end shield. To achieve pure radial offset without inducing a net moment on the shaft, diametrically opposed screw pairs were adjusted in a coordinated manner (one advanced while the other retracted). The resulting eccentricity was measured using a dial indicator, calibrated such that one full screw rotation produced a 0.250 mm radial displacement. To simulate progressive static eccentricity and facilitate comparative analysis of fault-induced features under different load conditions, two severity levels were implemented—corresponding to radial offsets of 0.125 mm and 0.250 mm.

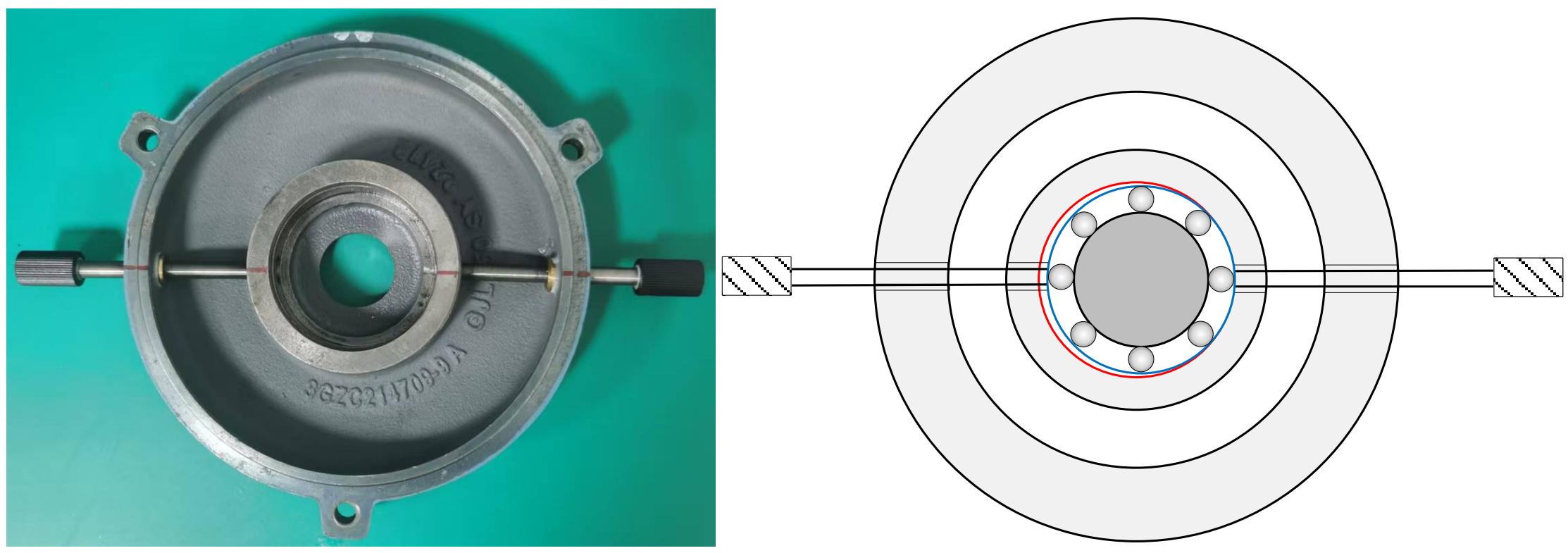

Fig. 4. Static-eccentricity setting screws

Rotor unbalance was induced by attaching an imbalance mass to the rotor shaft, as shown by the dashed outline in Figure 5.

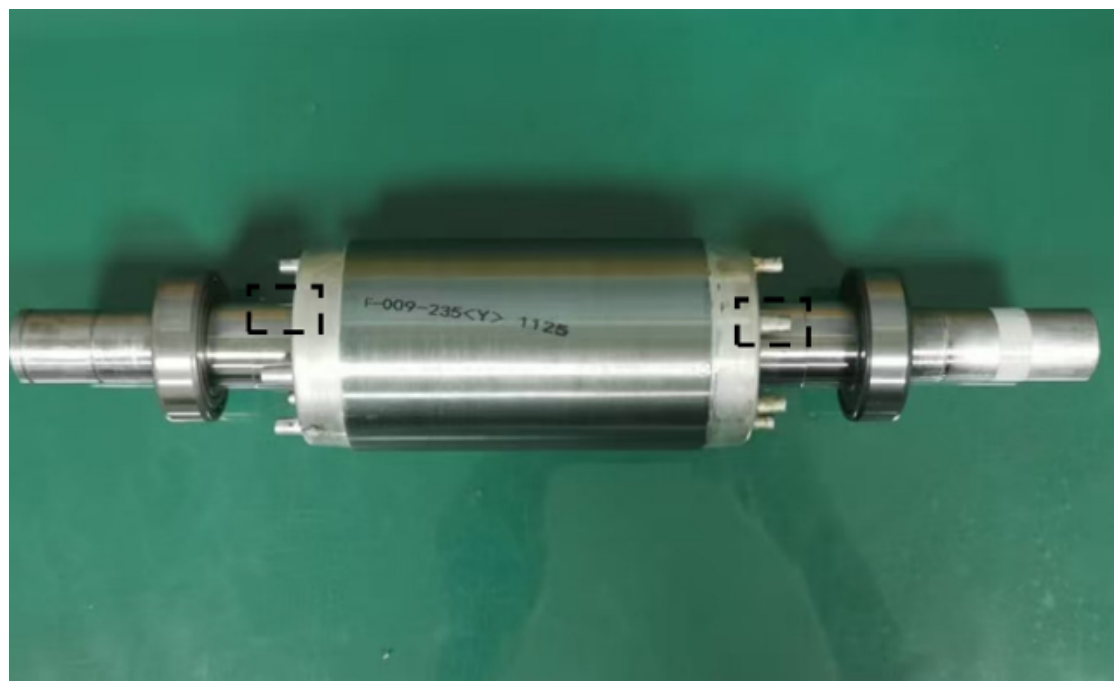

Fig. 5. Dynamic unbalance

Broken rotor bar fault was simulated by deliberately removing three consecutive rotor bars, as depicted in Figure 6. To compensate for the resultant mass loss and avoid introducing mechanical imbalance, counterweights were added to the end ring. This rebalancing ensured that the broken bar fault remained the sole electrical anomaly, thereby preventing contamination of subsequent analysis by other defect types.

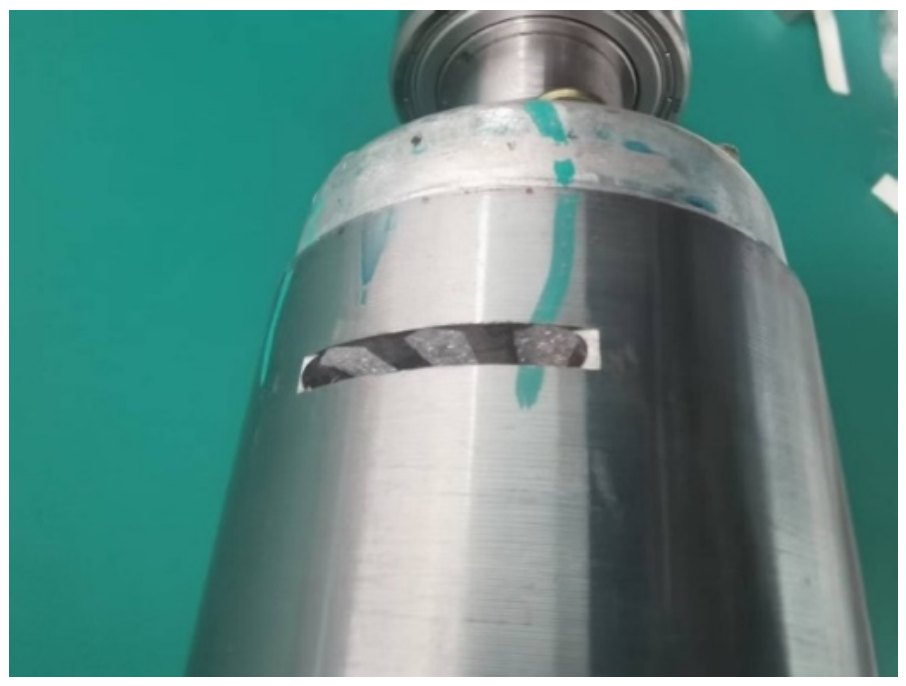

Fig. 6 Broken rotor bar.

Bearing faults were systematically categorized into three types: outer raceway, inner raceway, and rolling element damage. The dataset is limited to single-point raceway faults, as field data indicates simultaneous damage to both inner and outer races is rare. Consequently, defects were introduced exclusively on either the inner or the outer raceway, ensuring unambiguous attribution of characteristic frequencies, specifically the Ball Pass Frequency Inner race (BPFI) and the Ball Pass Frequency Outer race (BPFO).

All experiments employed SKF 6205 deep-groove ball bearings. Raceway notches were fabricated using a laser cutting process with dimensional tolerances of ±0.01 mm. Figure 7 shows an outer raceway defect, with Table 6 summarizing the width and depth dimensions corresponding to mild and severe damage levels.

Table 6 Raceway fault parameters

| Fault parameter of bearing | Light | High |
|---|---|---|
| Inner fault width | 0.2 mm | 0.6 mm |
| Inner fault depth | 0.5 mm | 0.5 mm |
| Outer fault width | 0.2 mm | 0.6 mm |
| Outer fault depth | 0.5 mm | 0.5 mm |

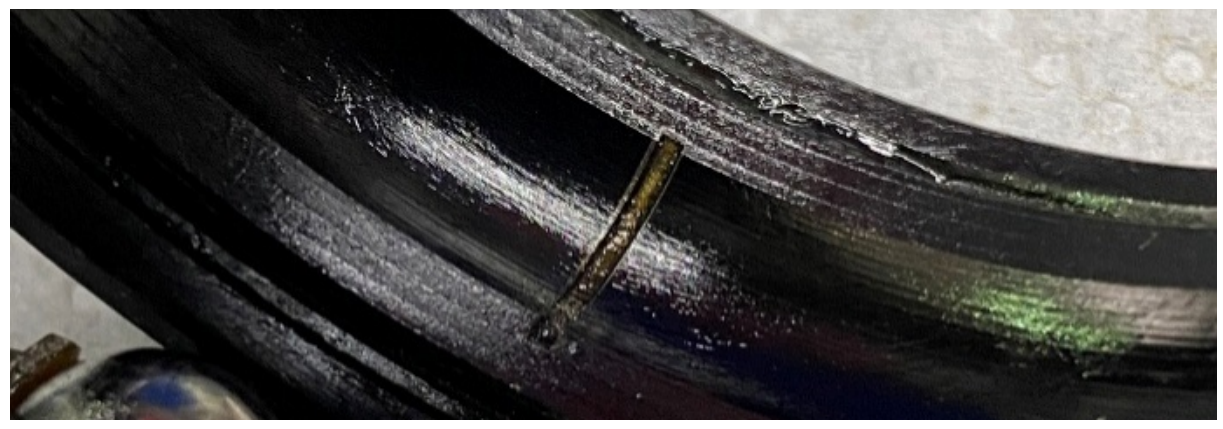

Fig. 7. outer ring raceway damage.

Rolling element damage, characterized by the Ball Spin Frequency (BSF), was implemented by introducing a 360° circumferential pitting on a single ball. The uniform distribution of the pits ensures contact with the raceways during every ball rotation (Figure 8).

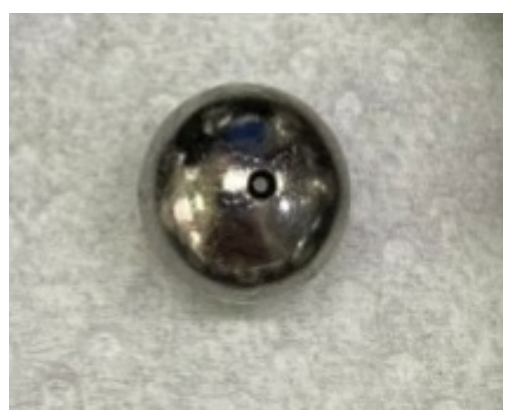

Fig. 8. Rolling element damage.

Finally, Table 7 lists the dimensionless fault frequency coefficients, defined as the ratio of the actual fault frequency to the shaft rotational frequency.

Table 7 Relative bearing damage frequencies

| Bearing | BPFO | BPFI | BSF |
|---|---|---|---|
| SKF 6205 2ZC3 | 3.585 | 5.415 | 2.357 |

To simulate a bent shaft condition, the motor rotor assembly was subjected to a four-point bending test, resulting in a permanent shaft bend of 0.2 mm. Voltage unbalance was simulated by connecting an external resistor in series with one phase of the motor power supply. Rotating a control knob increased the series resistance: when the resistor was bypassed (knob at minimum position), the phase voltage equaled the rated supply voltage; at the maximum knob position, the reduced current in that phase caused a measurable voltage drop at the motor terminals, creating voltage unbalance. Two calibrated knob positions, corresponding to 4% and 8% voltage unbalance levels, were used to generate different fault severities.

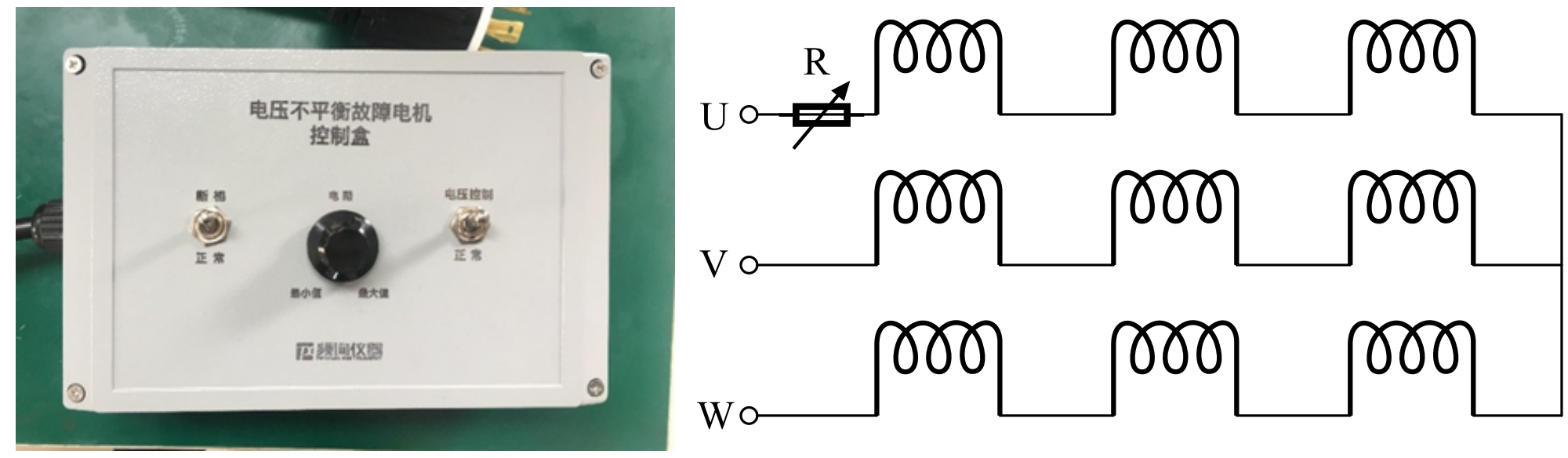

Fig. 9. Voltage unbalance fault controller

To further investigate the electromechanical coupling between motor-side electromagnetic anomalies and mechanical bearing degradation, the dataset incorporates multiple compound-fault scenarios in which a controlled bearing defect is introduced concurrently with another representative fault source. Specifically, a localized outer-race bearing defect is combined with (i) a fixed static radial offset to emulate static eccentricity, enabling analysis of eccentricity–bearing cross-effects; (ii) rotor unbalance to study interaction mechanisms between mass-imbalance–induced excitation and bearing deterioration; (iii) broken rotor bars to couple an electrical asymmetry with mechanical bearing damage; and (iv) an inter-turn winding short circuit to quantify electromagnetic–mechanical cross-modulation, where uneven phase magnetomotive forces and localized heating induce fault-related magnetic force fluctuations that may further interact with bearing-surface damage. Across these conditions, the two fault sources are introduced independently and then applied simultaneously, providing a systematic basis for characterizing coupled signatures and cross-modulation in compound electromechanical faults.

## USAGE NOTES

The dataset supports multiple benchmark task formulations. Representative examples and recommended practices are outlined below to facilitate the establishment of reproducible experimental settings under varying research objectives. In-condition diagnosis can be defined by partitioning the training, validation, and test sets within a single set of operating conditions, maintaining consistent fault-class proportions across splits. To prevent information leakage, it is advisable to first split at the recording-run level before applying windowing or segmentation, ensuring that samples from the same run do not appear in both training and testing sets. Unknown-condition evaluation can be instantiated by reserving one or more operating-condition groups as the test set, with training and model selection limited to the remaining known-condition groups. The validation set should be drawn exclusively from the known-condition pool. It is further recommended to report both overall metrics and condition-grouped metrics to evaluate robustness under operating-condition shifts. Transitional-condition robustness can be assessed by training on steady-state condition groups and testing on transitional or time-varying regimes. As with other settings, run-level splitting prior to segmentation is essential. Time-resolved or segment-level metrics may be additionally employed to capture performance degradation under nonstationary conditions.

Beyond task partitioning, the adoption of a consistent signal-processing pipeline enhances comparability and reproducibility. Sample construction can employ fixed-length sliding windows, with window length selected based on the target feature scale and an appropriate overlap ratio. Preprocessing may include mean removal or detrending, and high-pass or band-pass filtering should be applied according to the sampling bandwidth to mitigate drift and noise. Identical filtering parameters must be used across training and testing. Channel-wise normalization is recommended, using training-set statistics for standardization or amplitude normalization, and applying the same transformation to the validation and test sets to avoid test data leakage. These guidelines are recommended, not prescriptive, and may be adapted to specific research objectives, provided that partitioning remains leakage-free and preprocessing remains consistent.

## LIMITATIONS

Despite the comprehensive coverage of fault types and operating conditions, the dataset is limited to single-point bearing faults and does not include distributed or progressive fault patterns that may occur in long-term service. Additionally, the fault severity levels are discretely defined and may not fully represent the continuous degradation process observed in actual industrial settings. The experiments were conducted under controlled laboratory conditions with minimal electromagnetic interference, which may differ from the complex noise environments encountered in real-world applications.

## ETHICS STATEMENT

*The authors have read and follow the ethical requirements for publication in Data in Brief and confirming that the current work does not involve human subjects, animal experiments, or any data collected from social media platforms.*

## CRediT AUTHOR STATEMENT

**Shijin Chen:** Conceptualization, Methodology, Writing –original draft

**Zeyi Liu:** Conceptualization, Methodology, Writing –review & editing

**Chenyang Li:** Data curation, Investigation, Validation

**Dongliang Zou:** Conceptualization, Resources, Project administration

**Xiao He:** Conceptualization, Formal analysis, Supervision

**Donghua Zhou:** Supervision, Formal analysis

## ACKNOWLEDGEMENTS

*S. Chen, Z. Liu, C. Li contributed equally to this paper. This work was supported by National Natural Science Foundation of China under grants 624B2087, 62525308, 62473223, and 52172323, in part by the China MCC5 Group PhD innovation project under grant MCC5PHD20231107, in part by Beijing Natural Science Foundation under grant L241016.*

## DECLARATION OF COMPETING INTERESTS

The authors declare that they have no known competing financial interests or personal relationships that could have appeared to influence the work reported in this paper.